\documentclass[10pt,journal,twocolumn,final]{IEEEtran}
\usepackage{amssymb,amsmath,amsthm,amsfonts}
\usepackage{graphicx}
\usepackage{amsmath}
\usepackage{subfigure}
\usepackage{color}
\usepackage{epstopdf}
\usepackage{cite}
\usepackage{bbm}
\usepackage{algorithm}
\usepackage{algpseudocode}
\usepackage{array}
\usepackage[utf8]{inputenc}
\usepackage{cases}

\begin{document}

\title{Risk-Sensitive Task Fetching and Offloading for Vehicular Edge Computing}

\author{
\IEEEauthorblockN{
\begin{center}Sadeep Batewela, \IEEEmembership{Student Member, IEEE}, Chen-Feng Liu, \IEEEmembership{Student Member, IEEE},\\Mehdi Bennis, \IEEEmembership{Senior Member, IEEE}, Himal A. Suraweera, \IEEEmembership{Senior Member, IEEE},\\and Choong Seon Hong, \IEEEmembership{Senior Member, IEEE}
\end{center}}
\thanks{This work was supported in part by the Academy of Finland project CARMA, in part by the Academy
of Finland project MISSION, in part by the Academy of Finland project SMARTER, in part by the INFOTECH project NOOR, in part by the Kvantum Institute strategic
project SAFARI, and in part by the Nokia Foundation.}
\thanks{S. Batewela  is with Nokia Networks, 33100 Tampere, Finland (e-mail: sadeep.batewela\_vidanelage@nokia.com).}
\thanks{C.-F. Liu is with the Centre for Wireless Communications, University of Oulu,  90014 Oulu, Finland (e-mail: chen-feng.liu@oulu.fi).}
\thanks{M. Bennis is with the Centre for Wireless Communications, University of Oulu,  90014 Oulu, Finland, and also with the Department of Computer Science and Engineering, Kyung Hee University, Yongin 17104, South Korea (e-mail: mehdi.bennis@oulu.fi).}
\thanks{H. A. Suraweera is with the Department of Electrical and Electronic Engineering, University of Peradeniya, Peradeniya 20400, Sri Lanka  (e-mail: himal@ee.pdn.ac.lk).}
\thanks{C. S. Hong is with the Department of Computer Science and Engineering, Kyung Hee University, Yongin 17104, South Korea  (e-mail: cshong@khu.ac.kr).}
}

\maketitle

\begin{abstract}
This letter studies an ultra-reliable low latency communication problem focusing on a vehicular edge computing network in which vehicles either fetch and synthesize images recorded by surveillance cameras or acquire the synthesized image from an edge computing server. The notion of \emph{risk-sensitive} in financial mathematics is leveraged to define a reliability measure, and the studied problem is formulated as a risk minimization problem for each vehicle's end-to-end (E2E) task fetching and offloading delays. Specifically, by resorting to a joint utility and policy estimation-based learning algorithm, a distributed risk-sensitive solution for task fetching and offloading is proposed. Simulation results show that our proposed solution achieves performance improvements up to 40\% variance reduction and steeper distribution tail of the E2E delay over an averaged-based baseline.

\end{abstract}

\section{Introduction}
\label{sec:intro}

\IEEEPARstart{B}{y applying} the mobile edge computing (MEC) architecture \cite{fog_survey} to vehicular networks,
vehicular edge computing (VEC) \cite{Ning_VehMag19} provides computation service for vehicles to execute time-critical applications.
The main problem is that the latency exacerbates when all vehicles intend to access the MEC server's computational resources.
To this end, some works have  studied the  offloading and resource allocation mechanisms in various VEC scenarios \cite{Zhang_ICC17,Xu_GC18,Dai_IoTJ19}.
However, the majority of the VEC literature  aimed at reducing the latency,
whereas reliably delivering messages for traffic safety maintenance is critical in VEC networks.
To enable ultra-reliable low latency communication (URLLC), improving the delay performance on average  is not sufficient \cite{Mehdi_Proc18} but requires further breaking down the statistics of delay, i.e., mean, variance, skewness, etc. 
To this end, we resort to the \emph{risk} measure, a notion adopted in financial mathematics \cite{Risk}, for wireless communication, where risk is synonymous with losing valuable information due to the randomness of wireless transmission. For example, the stochastic channel quality incurs delay variation, and a higher variance in the delay can result in an urgent-message loss, making traffic safety at stake.

In this letter, we consider a VEC scenario in which  computation tasks are remotely generated  at  surveillance cameras.  Next, the vehicle decides either to fetch the tasks from the cameras for local computation or to obtain the computed results from the MEC server. In the latter case, tasks are offloaded from the cameras and computed at the server.
Motivated by the aforementioned shortcomings, the studied problem is cast as a risk minimization problem for each vehicle's end-to-end (E2E) delay.
Moreover, the variance and higher-order statistics of the E2E delay are taken into account in the risk-minimization problem. 
Since all vehicles' delays are coupled together in the considered system, each vehicle requires other vehicles' wireless channel information, which is not available in practice, to solve the  problem. To address this issue,
we leverage upon a joint utility and policy estimation-based learning framework \cite{Sun_Survey19,Sumudu_regret} and propose a distributed risk-sensitive approach for task fetching and offloading.
Numerical results show that in contrast to the average-based system design, our proposed solution achieves lower variance and steeper distribution tail of the E2E delay albeit higher average performance.
The superiority of the proposed approach in denser networks is validated through comparison with several baseline schemes.

\section{System Model}
\label{Sec:system}

\begin{figure}[t]
\center
\includegraphics[width=\columnwidth]{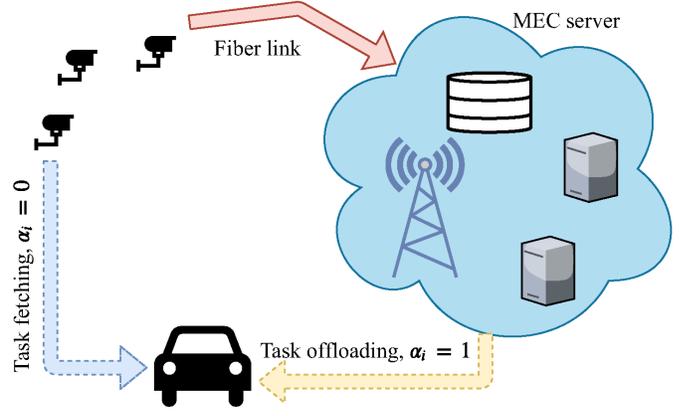}
\caption{Schematic illustration of the task-fetching and offloading decision in a vehicular setting.}
\label{model}
%\vspace{-0.5cm}
\end{figure}

We consider an urban VEC network around the road intersection, where a set $\mathcal{V}$ of $V$ vehicular user equipments (VUEs) locate on the intersecting roads. A set $\mathcal{C}$ of $C$ cameras are installed at the intersection to record images for aiding traffic safety. We assume that each camera  has a limited angle of view, and all cameras' images need to be synthesized to have the full view of the intersection. Since VUEs are at different locations, we simply assume that each VUE requires a dedicated synthesized image which is different from the other VUEs' required synthesized images. The VUE has its own computational capability to synthesize the images that are wirelessly fetched from cameras. Moreover, due to VUEs' limited computation capabilities, an MEC server, connected with all cameras via fiber, is deployed to provide the computational service for the VUEs. In this regard,  images are first delivered to the MEC server through the fiber links. After synthesizing, the MEC server wirelessly sends the synthesized images to the VUEs.
Let us further denote each VUE $i$'s task-fetching and offloading decision as $\alpha_i\in\{0,1\},\forall\,i\in\mathcal{V}$, in which the VUE fetches images and synthesize them locally if $\alpha_i=0$. Otherwise, synthesizing is executed at the MEC server. 
The task-fetching and offloading decision in the considered VEC system is schematically shown in Fig.~\ref{model}.

When $\alpha_i=0$, the VUE first fetches the images from the cameras. We assume that each camera has a dedicated bandwidth $W_{\rm c}$ to broadcast its image to all task-fetching VUEs, denoted by  $\mathcal{V}_{\rm f}=\{i\in\mathcal{V}|\,\alpha_i=0\}$. In order to ensure that all VUEs in $\mathcal{V}_{\rm f}$ can correctly receive the same image over the full bandwidth $W_{\rm c}$, each camera $j\in\mathcal{C}$ broadcasts its data with the rate $R_{j}=W_{\rm c}\log_2\big(1+\frac{P_{\rm c}h_{j,\min}}{W_{\rm c}N_0}\big)$, considering the minimal channel gain among all corresponding camera-VUE links, i.e., $h_{j,\min}=\underset{i\in\mathcal{V}_{\rm f}}{\min}\,h_{ji}$. Here, $h_{ji}$ is the channel gain, including path loss and channel fading, between camera $j$ and VUE $i$. We further assume that all wireless channel gains belong to the finite sets such that channels stay static\footnote{For example, by quantizing the continuous channel fading values in consecutive coherence time blocks into the same level such that the corresponding transmission rate is sustainable.} during task fetching and offloading.
$P_{\rm c}$ is the camera's transmit power which is identical for all cameras, and $N_0$ is the power spectral density of the additive white Gaussian noise. Then given the camera's image size $A$, the transmission delay from camera $j$ is $T^{\rm Tx}_{j}=A/R_{j}$. Since the VUE $i$ has to wait till it receives all cameras' images before synthesizing, the net transmission delay for each VUE is $T^{\rm Fet}=\underset{j\in\mathcal{C}}{\max}\,T^{\rm Tx}_{j}$. Here, we note that the transmission delay $T^{\rm Fet}$ is affected by all the task-fetching VUEs' wireless channels as per $h_{j,\min}=\underset{i\in\mathcal{V}_{\rm f}}{\min}\,h_{ji}$. Assuming that all camera's images are in the same size, we calculate the computation delay for image synthesizing at VUE $i$ as $T^{\rm Comp}_{i}=CAL/f_{i}$ in which $L$ and $f_i$ (in the unit of cycle/sec) are, respectively, the required central processing unit (CPU) cycles per bit for computation, i.e., the processing density, and VUE $i$'s local CPU-cycle frequency \cite{fog_survey}. When $\alpha_{i}=0$, VUE $i$'s E2E delay $T^{\rm E2E}_{i}$ includes the transmission delay and local computation delay, i.e., $T^{\rm E2E}_{i}=T^{\rm Fet}+T^{\rm Comp}_{i}$.

For $\alpha_i=1$, let us first analogously denote the set of VUEs with the task-offloading decision as $\mathcal{V}_{\rm o}=\{i\in\mathcal{V}|\,\alpha_i=1\}$. Additionally, since all cameras deliver their images to the MEC server through fiber links, we neglect the transmission delays from cameras to the MEC server. After obtaining these images, the MEC server equally divides its  total CPU-cycle frequency $f_{\rm s}$ into $|\mathcal{V}_{\rm o}|$ parts in order to simultaneously synthesize the $|\mathcal{V}_{\rm o}|$ VUEs' dedicated images. Thus, the computational delay at the MEC server is expressed as $T^{\rm Comp}_{\rm s}= C A L\lvert\mathcal{V}_{\rm o}\rvert/f_{\rm s}$. Subsequently, the MEC server orthogonally sends the synthesized image to each VUE $i\in\mathcal{V}_{\rm o}$ with the equally allocated bandwidth $W_{\rm s}/|\mathcal{V}_{\rm o}|$. The allocated transmit power for VUE $i$ is denoted by $P_i$ with the constraint
\begin{equation}
\sum\limits_{i\in\mathcal{V}_{\rm o}}P_{i}=P_{\max}\mbox{ and }P_{i}\geq 0,~\forall\,i\in\mathcal{V}_{\rm o},\label{Eq: Power}
\end{equation}
where $P_{\max}$ is the MEC server's transmit power budget. The power allocation mechanism will be detailed in Section \ref{Sec: Server's power}.
Accordingly, the corresponding downlink (DL) rate (from the MEC server) to the VUE $i$ is expressed as $R_{i}=\frac{W_{\rm s}}{\lvert \mathcal{V}_{\rm o}\rvert}\log_2\big(1+\frac{P_{i}h_{{\rm s}i}\lvert \mathcal{V}_{\rm o}\rvert}{W_{\rm s}N_0}\big)$, where $h_{{\rm s}i}$ is the DL channel gain from the MEC server to VUE $i$. Then given the synthesized image size $B$, the DL transmission delay of the VUE $i$ is $T^{\rm DL}_{ i}=B/R_{i}$. When $\alpha_{i}=1$, VUE $i$'s E2E delay includes the computation delay at the MEC server and the DL transmission delay, i.e., $T^{\rm E2E}_{i}=T^{\rm Comp}_{\rm s}+T^{\rm DL}_{ i}$. In summary, VUE $i$'s E2E latency is expressed as
\begin{align}
T^{\rm E2E}_{i}=
\begin{cases}
\underset{j\in\mathcal{C}}{\max}\left\{\frac{A}{W_{\rm c}\log_2\big(1+\frac{P_{\rm c}h_{j,\min}}{W_{\rm c}N_0}\big)}\right\}+\frac{CAL}{f_{i}},
\\\hspace{11.5em}\mbox{when~}\alpha_i=0,
\\\frac{CAL\lvert\mathcal{V}_{\rm o}\rvert }{f_{\rm s}}+\frac{B\lvert \mathcal{V}_{\rm o}\rvert}{W_{\rm s}\log_2\big(1+\frac{P_{i}h_{{\rm s}i}\lvert \mathcal{V}_{\rm o}\rvert}{W_{\rm s}N_0}\big)},
\\\hspace{11.5em}\mbox{when~}\alpha_i=1.
\end{cases}\label{Eq: E2E_offloading}
\end{align}
Furthermore, we note that the E2E delay in \eqref{Eq: E2E_offloading} is random.

\section{Problem Formulation}\label{Sec: Problem}

\subsection{Risk Minimization for the VUE's End-to-End Delay}

As mentioned in Section \ref{sec:intro}, risk is considered as our concerned reliability metric.
In this regard, we focus on the entropic risk measure\footnote{Because entropic risk measure directly incorporates the statistics of the delay, i.e., mean, variance, skewness, etc.} $\frac{1}{\rho}\ln(\mathbb{E}[\exp(\rho T^{\rm E2E}_{i})])$ and formulate a risk minimization problem for each VUE $i\in\mathcal{V}$ as follows:
\begin{subequations}\label{Eq: entropy_VUE_problem}
\begin{IEEEeqnarray}{cll}
\underset{\Pr(\alpha_i|\mathbf{h}_{i})\geq 0}{\mbox{minimize}}&~~\frac{1}{\rho}\ln(\mathbb{E}[\exp(\rho T^{\rm E2E}_{i})])&\label{Eq: entropy_VUE_problem_1}
\\\mbox{subject to}&~~\sum\limits_{\alpha_i\in\{0,1\}}\Pr(\alpha_i|\mathbf{h}_{i})=1,&~~\forall\,\mathbf{h}_{i}\in\mathcal{H}_i,\label{Eq: entropy_VUE_problem_2}
\end{IEEEeqnarray}
\end{subequations}
where $\mathbb{E}[\cdot]$ is the expectation operator, and $\mathbf{h}_{i}=[h_{ji},h_{{\rm s}i}:j\in\mathcal{C}]\in\mathcal{H}_i,\forall\,i\in\mathcal{V}$, denotes VUE $i$'s channel vector with a finite set $\mathcal{H}_i$. By taking the Maclaurin series expansion, we can express the objective function \eqref{Eq: entropy_VUE_problem_1} as $\frac{1}{\rho}\ln(\mathbb{E}[\exp(\rho T^{\rm E2E}_{i})])=\mathbb{E}[T^{\rm E2E}_{i}]+\frac{\rho}{2!}{\rm Var}(T^{\rm E2E}_{i})+\frac{\rho^2\mu_3}{3!}+\cdots$ in which $\mu_3=\mathbb{E}[(T^{\rm E2E}_{i}-\mathbb{E}[T^{\rm E2E}_{i}])^3]$ is the third central moment of $T^{\rm E2E}_{i}$, and the skewness of $T^{\rm E2E}_{i}$ is equal to $\frac{\mu_3}{({\rm Var}(T^{\rm E2E}_{i}))^{3/2}}$. In other words, the mean, variance, skewness, and other higher-order statistics of the E2E delay are taken into account in the objective. Moreover, the risk-sensitivity parameter $\rho>0$ reflects the weight of higher-order statistics in the formulated risk minimization problem. Owing to the randomness of $\mathbf{h}_{i}$ and other unavailable information, e.g., the other VUEs' channels, each VUE $i$ focuses on the probabilistic policy $\Pr(\alpha_i|\mathbf{h}_{i})$ of the task-fetching and offloading decision in problem \eqref{Eq: entropy_VUE_problem}. Since $\frac{1}{\rho}\ln(\cdot)$ in \eqref{Eq: entropy_VUE_problem_1} is a monotonically increasing function, we can directly remove it and focus on an equivalent  problem
\begin{IEEEeqnarray}{cl}\label{Eq: VUE_problem_1}
\underset{\Pr(\alpha_i|\mathbf{h}_{i})\geq 0}{\mbox{minimize}}&~\mathbb{E}[\exp(\rho T^{\rm E2E}_{i})]~~
\mbox{subject to}~\eqref{Eq: entropy_VUE_problem_2}.
\end{IEEEeqnarray}
Analogously, by taking the Maclaurin series expansion, i.e.,
$\mathbb{E}[\exp(\rho T^{\rm E2E}_{i})]=1+\rho\mathbb{E}[T^{\rm E2E}_{i}]+\frac{\rho^2}{2!}\mathbb{E}[(T^{\rm E2E}_{i})^2]+\frac{\rho^3}{3!}\mathbb{E}[(T^{\rm E2E}_{i})^3]+\cdots$, we can see that
the higher-order moments/statistics are still considered in the objective function of the equivalent problem.

\subsection{Transmit Power Allocation at the MEC Server}\label{Sec: Server's power}

Referring to the motivation of considering \eqref{Eq: VUE_problem_1}, we formulate the MEC server's power allocation problem as
\begin{IEEEeqnarray}{c}
\underset{P_{i}}{\mbox{minimize}}~\sum\limits_{i\in\mathcal{V}_{\rm o}}\exp(\rho T^{\rm DL}_{i})
~~\mbox{subject to}~\eqref{Eq: Power}.\label{Eq: Server_problem}
\end{IEEEeqnarray}
In the objective, we consider the DL transmission delay since the allocated transmit power only affects this delay. Following the results in \cite[Eq.~(3.10)]{Boyd}, we can verify that $\exp(\rho T^{\rm DL}_{i})= 
\exp\Big(\frac{\rho B\lvert \mathcal{V}_{\rm o}\rvert}{W_{\rm s}\log_2\big(1+\frac{P_{i}h_{{\rm s}i}\lvert \mathcal{V}_{\rm o}\rvert}{W_{\rm s}N_0}\big)}\Big)
$ is a convex function with respect to $P_i$. Hence, we directly apply the Karush--Kuhn--Tucker (KKT) conditions and derive the following optimal solution. For the DL transmission, the MEC server allocates  
the transmit power $P^{*}_{i}>0,\forall\, i\in\mathcal{V}_{\rm o}$, which satisfies
$\frac{\theta\kappa_{i}\exp\big(\frac{\theta}{\ln(1+P^{*}_{i}\kappa_{i})}\big)}{(1+P^{*}_{i}\kappa_{i})[\ln(1+P^{*}_{i}\kappa_{i})]^2}=\nu$
with
$\theta=\frac{\rho B\lvert \mathcal{V}_{\rm o}\rvert \ln 2}{W_{\rm s}}$ and $\kappa_{i}=\frac{h_{{\rm s}i}\lvert \mathcal{V}_{\rm o}\rvert}{W_{\rm s}N_0}$.
Here, $\nu$ is chosen such that
$\sum_{i\in\mathcal{V}_{\rm o}}P_{i}^{*}=P_{\max}$.

\section{Distributed Risk-Sensitive Approach for Task Fetching and Offloading}\label{Sec: approach}

Note that each VUE $i$'s objective $\mathbb{E}[\exp(\rho T^{\rm E2E}_{i})]$ in problem \eqref{Eq: VUE_problem_1} is affected by not only its own policy of the task fetching and offloading decision but also the other VUEs' as per \eqref{Eq: Power} and \eqref{Eq: E2E_offloading}. Hence,
the studied risk minimization for all VUEs is a multi-agent problem, where the stable\footnote{While all the other VUEs follow their stable policies, the VUE deviating from the stable policy cannot further decrease its objective $\mathbb{E}[\exp(\rho T^{\rm E2E}_{i})]$.} policy of \eqref{Eq: VUE_problem_1} is desired at each VUE. To solve \eqref{Eq: VUE_problem_1},  each VUE $i$ requires the full information of $|\mathcal{V}_{\rm o}|$, $\mathcal{V}_{\rm f}$, and $\mathbf{h}_{i'},\forall\,i'\in\mathcal{V}_{\rm f}\setminus i$, as per \eqref{Eq: E2E_offloading}. In other words, the VUE needs to know all possible values of $\exp(\rho T^{\rm E2E}_{i})$ in the objective function. However, the required information is cumbersome to fetch in practice. 
To address this issue while achieving the stable policy, we consider a joint utility and policy estimation-based learning framework, i.e., the distributed no-regret learning algorithm \cite{Sumudu_regret},  which iteratively estimates the average impact of the aforementioned unavailable information on $\exp(\rho T^{\rm E2E}_{i})$ and
further obtain a task-fetching and offloading solution to problem \eqref{Eq: VUE_problem_1}.
The main steps of the distributed no-regret learning algorithm are outlined as follows:
\begin{itemize}
\item
Each VUE $i$ is given an initial policy $\Pr(\alpha_i|\mathbf{h}_{i})$.
\item
In each iteration $t$, VUE $i$ observes the independent and identically distributed (i.i.d.) realization $\mathbf{h}_i(t)$  and then makes a decision $\alpha_i(t)$ based on its policy $\Pr(\alpha_i|\mathbf{h}_{i})$.
\item
Subsequently, cameras and the MEC server are implicitly informed about the VUE compositions of  $\mathcal{V}_{\rm f}(t)$ and $\mathcal{V}_{\rm o}(t)$, respectively, and send the images to the corresponding VUEs. Once the synthesized image is available, each VUE $i$ can observe its E2E delay $T^{\rm E2E}_{i}(t)$ and calculate the utility  $-\exp(\rho T^{\rm E2E}_{i}(t))$. 
\item
Each VUE calculates the average utilities over all past iterations by 1) always making the same decision $\alpha_i$; and 2) following its iteration-variant policy $\Pr(\alpha_i|\mathbf{h}_{i})$. The \emph{regret} of not insisting on a specific decision is the difference between the average utilities of  1) and 2). 
\item
VUE $i$ iteratively updates its policy such that the likelihood of making a specific decision is proportional to the corresponding regret calculated over the past iterations.
\end{itemize}
Let us explain the details in the remainder of this section.
Firstly, we denote all unavailable information of VUE $i$ as  $\boldsymbol{\omega}_i$. Since the objective function of problem \eqref{Eq: VUE_problem_1} is affected by $\alpha_i$, $\mathbf{h}_i$, and $\boldsymbol{\omega}_i$, we define VUE $i$'s utility in the  $t$-th iteration as $u_i(t)=u_i(\alpha_i(t),\mathbf{h}_i(t),\boldsymbol{\omega}_i(t))=-\exp(\rho T^{\rm E2E}_{i}(t))$.
Additionally, for notational clarity, we use $\alpha^{m}_i$ and $\mathbf{h}_i^{l}$ to express the specific realizations of $\alpha_i(t)$ and $\mathbf{h}_i(t)$, respectively. 
The  \emph{state} realizations of  $\mathbf{h}_i(t)$ in all iterations are i.i.d., whereas the \emph{action} realization of $\mathbf{\alpha}_i(t)$ is generated based on some iteration-variant distribution $\Pr(\alpha_i|\mathbf{h}_{i})={\pi}_{\mathbf{h}_i}(t-1;\alpha_i)$.
Subsequently, considering the state realization $\mathbf{h}^{l}_i$, the conditional \emph{regret} of the action $\alpha^{m}_i$ in iteration $t$ is defined as
\begin{multline}\label{Eq: regret}
r_{\mathbf{h}_i^{l}}(t;\alpha^{m}_i)=\frac{1}{\sum_{\tau=1}^{t}\mathbbm{1}_{\{\mathbf{h}_i(\tau)=\mathbf{h}^{l}_i\}}}
\\\times\sum\limits_{\tau=1}^{t} \big[u_i(\alpha^{m}_i,\mathbf{h}_i^l,\boldsymbol{\omega}_i(\tau))-u_i(\tau)\big]\times\mathbbm{1}_{\{\mathbf{h}_i(\tau)=\mathbf{h}^{l}_i\}},
\end{multline}
where  $\mathbbm{1}_{\{\cdot\}}$ represents the indicator function. In \eqref{Eq: regret}, we calculate the average utility over the past iteration instants with  $\mathbf{h}_i(\tau)=\mathbf{h}^{l}_i$ and constantly changing $\alpha_i(\tau)$, and the average utility of fixing the specific action  $\alpha^{m}_i$ over these instants.
The conditional regret $r_{\mathbf{h}_i^{l}}(t;\alpha^{m}_i)$ is interpreted as the average utility improvement by fixing the decision $\alpha^{m}_i$ in the iteration instants with $\mathbf{h}_i(\tau)=\mathbf{h}^{l}_i$. However, since the realization of $\boldsymbol{\omega}_i(\tau)$ is unknown, it is not feasible to calculate  $u_i(\alpha^{m}_i,\mathbf{h}_i^l,\boldsymbol{\omega}_i(\tau))$ and the conditional regret \eqref{Eq: regret}.
To deal with this issue, the VUE estimates the expected utility with respect to $\boldsymbol{\omega}_i$ given $\alpha^{m}_i$ and $\mathbf{h}^l_i$, i.e., $\mathbb{E}_{\boldsymbol{\omega}_i}[u_i(\alpha^{m}_i,\mathbf{h}^l_i)]$. To this end, the expected utility $\mathbb{E}_{\boldsymbol{\omega}_i}[u_i(\alpha^{m}_i,\mathbf{h}^l_i)]$ and conditional regret $r_{\mathbf{h}_i^{l}}(t;\alpha^{m}_i)$ are estimated, in iteration $t$, as per
\begin{align}
\hspace{-0.7em}\hat{u}_{\mathbf{h}_i^{l}}(t;\alpha^{m}_i)&=\hat{u}_{\mathbf{h}_i^{l}}(t-1;\alpha^{m}_i)+\eta_{u}(t)\times\mathbbm{1}_{\{\mathbf{h}_i(t)=\mathbf{h}^{l}_i\}}\notag
\\&\quad\times\mathbbm{1}_{\{\alpha_i(t)=\alpha^{m}_i\}}\times\big[u(t)-\hat{u}_{\mathbf{h}_i^{l}}(t-1;\alpha^{m}_i)\big],\label{Eq: Learning update-1}
\\\hspace{-0.7em}\hat{r}_{\mathbf{h}_i^{l}}(t;\alpha^{m}_i)&=\hat{r}_{\mathbf{h}_i^{l}}(t-1;\alpha^{m}_i)+\eta_{r}(t)\times\mathbbm{1}_{\{\mathbf{h}_i(t)=\mathbf{h}^{l}_i\}}\notag
\\&\quad\times\big[\hat{u}_{\mathbf{h}_i^{l}}(t;\alpha^{m}_i)-u(t)-\hat{r}_{\mathbf{h}_i^{l}}(t-1;\alpha^{m}_i)\big],\label{Eq: Learning update-2}
\end{align}
$\forall\,\alpha_i^{m}\in\{0,1\},\mathbf{h}_i^{l}\in\mathcal{H}_i$. Given all the estimated conditional regrets $\hat{r}_{\mathbf{h}_i^{l}}(t;\alpha^{m}_i)$ in the current iteration $t$, a straightforward solution to the task-fetching and offloading problem is making the decision (for each channel vector value $\mathbf{h}^l_i$) with the maximal regret. On the other hand, we also want to comprehensively explore the impacts of the randomness $\boldsymbol{\omega}_i$ on both $u_i(\alpha_i=0,\mathbf{h}^l_i)$ and $u_i(\alpha_i=1,\mathbf{h}^l_i)$. Incorporating these two concerns, the conditional probability distribution for the task-fetching and offloading decision in iteration $t$ can be modeled as \cite{Sumudu_regret}, $\forall\,\mathbf{h}_i^{l}\in\mathcal{H}_i$,
 $\beta(\hat{\mathbf{r}}_{\mathbf{h}_i^{l}}(t);\alpha_i)=\underset{\Pr(\alpha_i|\mathbf{h}_i^{l})}{\arg\max}\bigg\{\sum\limits_{\alpha^{m}_i=0}^{1}\big[\Pr(\alpha^{m}_i|\mathbf{h}_i^{l})\hat{r}_{\mathbf{h}_i^{l}}(t;\alpha^{m}_i)
-\frac{1}{\xi}\Pr(\alpha^{m}_i|\mathbf{h}_i^{l})\ln\big(\Pr(\alpha^{m}_i|\mathbf{h}_i^{l})\big)\big]\Big\}$
whose solution is given by
\begin{algorithm}[t]
  \caption{Distributed No-Regret Learning Algorithm}
  \begin{algorithmic}[1]
    \State Initialize $t=1$ and set initial values for $\hat{u}_{\mathbf{h}_i^{l}}(0;\alpha^{m}_i)$, $\hat{r}_{\mathbf{h}_i^{l}}(0;\alpha^{m}_i)$, ${\pi}_{\mathbf{h}_i^{l}}(0;\alpha^{m}_i)$, $\forall\,i\in\mathcal{V},\alpha^{m}_i\in\{0,1\},\mathbf{h}_i^{l}\in\mathcal{H}_i$.
    \Repeat
    \State Observing a realization $\mathbf{h}_i(t)=\mathbf{h}_i^{l}$, each VUE $i\in\mathcal{V}$  makes a decision $\alpha_i(t)=\alpha^{m}_i$ based on its $\Pr(\alpha_i(t)=\alpha^{m}_i|\mathbf{h}_i(t)=\mathbf{h}_i^{l})={\pi}_{\mathbf{h}_i^{l}}(t-1;\alpha_i^m)$.
      \State The MEC server allocates the transmit power for the VUE set $\mathcal{V}_{\rm o}(t)$ as per Section \ref{Sec: Server's power}.
       \State Each VUE $i$ observes its $T^{\rm E2E}_{i}(t)$ and locally updates  \eqref{Eq: Learning update-1}--\eqref{Eq: Learning update-3}.
      \State $t\leftarrow t+1$.
      \Until{Stopping criteria are satisfied.}
  \end{algorithmic}\label{Alg: distributed non-regret}
\end{algorithm}
\begin{equation}
\beta(\hat{\mathbf{r}}_{\mathbf{h}_i^{l}}(t);\alpha^{m}_i)=\frac{\exp\big(\xi \hat{r}^{+}_{\mathbf{h}_i^{l}}(t;\alpha^{m}_i)\big)}{\sum_{\tilde{\alpha}^{m}_i=0}^{1}\exp\big(\xi \hat{r}^{+}_{\mathbf{h}_i^{l}}(t;\tilde{\alpha}^{m}_i)\big)},\label{Eq: no-regret prob}
\end{equation}
$\forall\,\alpha_i^{m}\in\{0,1\}$, with $\hat{\mathbf{r}}_{\mathbf{h}_i^{l}}(t)=[\hat{r}_{\mathbf{h}_i^{l}}(t;0),\hat{r}_{\mathbf{h}_i^{l}}(t;1)]$ and $\hat{r}^{+}_{\mathbf{h}_i^{l}}(t;\alpha^{m}_i)=\max\{\hat{r}_{\mathbf{h}_i^{l}}(t;\alpha^{m}_i),0\}$.
Here, $\xi> 0$ is the parameter which trades off exploitation (maximizing the expected conditional regret) and exploration (maximizing information entropy).
Then combining the estimation rules in \eqref{Eq: Learning update-1} and \eqref{Eq: Learning update-2} with the distribution \eqref{Eq: no-regret prob}, we update the task-fetching and offloading policy for VUE $i$ as
\begin{align}
{\pi}_{\mathbf{h}_i^{l}}(t;\alpha^{m}_i)&={\pi}_{\mathbf{h}_i^{l}}(t-1;\alpha^{m}_i)+\eta_{\pi}(t)\times\mathbbm{1}_{\{\mathbf{h}_i(t)=\mathbf{h}^{l}_i\}}\notag
\\&\quad\times\big[\beta(\hat{\mathbf{r}}_{\mathbf{h}_i^{l}}(t);\alpha^{m}_i)-{\pi}_{\mathbf{h}_i^{l}}(t-1;\alpha^{m}_i)\big],\label{Eq: Learning update-3}
\end{align}
$\forall\,\alpha_i^{m}\in\{0,1\},\mathbf{h}_i^{l}\in\mathcal{H}_i$. Additionally, in \eqref{Eq: Learning update-1}, \eqref{Eq: Learning update-2}, and \eqref{Eq: Learning update-3}, the learning rates  $\eta_{u}(t)$, $\eta_{r}(t)$, and $\eta_{\pi}(t)$, which satisfy
$\lim\limits_{N\to\infty}\sum_{t=1}^{N}\eta_{u}(t)=\infty, \lim\limits_{N\to\infty}\sum_{t=1}^{N}\eta_{r}(t)=\infty,
\lim\limits_{N\to\infty}\sum_{t=1}^{N}\eta_{\pi}(t)=\infty,\lim\limits_{N\to\infty}\sum_{t=1}^{N}[\eta_{u}(t)]^2<\infty,
\lim\limits_{N\to\infty}\sum_{t=1}^{N}[\eta_{r}(t)]^2<\infty,\lim\limits_{N\to\infty}\sum_{t=1}^{N}[\eta_{\pi}(t)]^2<\infty,
\lim\limits_{t\to\infty}\frac{\eta_{r}(t)}{\eta_{u}(t)}=0$, and $\lim\limits_{t\to\infty}\frac{\eta_{\pi}(t)}{\eta_{r}(t)}=0$,
can be chosen by referring to $p$-series. In the next iteration $t+1$, the action realization of $\alpha_i(t+1)$ is generated based on ${\Pr}(\alpha_i(t+1)=\alpha^{m}_i|\mathbf{h}_i(t+1)=\mathbf{h}_i^{l})={\pi}_{\mathbf{h}_i^{l}}(t;\alpha^{m}_i)$. Finally, the converged distribution ${\pi}_{\mathbf{h}_i^{l}}(\infty;\alpha^{m}_i)$ provides a  stable solution to the studied problem \eqref{Eq: VUE_problem_1}.  The related proof of stability can be found in \cite{Sumudu_regret}.
The steps of the distributed no-regret learning algorithm are outlined in Algorithm \ref{Alg: distributed non-regret}. Moreover, given that $\mathbb{E}_{\boldsymbol{\omega}_i}[\exp(\rho T^{\rm E2E}_{i})|\alpha_i,\mathbf{h}_i],\forall\,\alpha_i,\mathbf{h}_i$, is known, problem \eqref{Eq: VUE_problem_1} is a linear programming problem which includes $2|\mathcal{H}_i|$ optimization variables and $3|\mathcal{H}_i|$ constraints. Here, $|\mathcal{H}_i|$ increases exponentially with the number $|\mathcal{C}|+1$. In Algorithm \ref{Alg: distributed non-regret} which addresses the unavailability dilemma of $\mathbb{E}_{\boldsymbol{\omega}_i}[\exp(\rho T^{\rm E2E}_{i})|\alpha_i,\mathbf{h}_i]$, $6|\mathcal{H}_i|$ memory elements  (for calculating \eqref{Eq: Learning update-1}, \eqref{Eq: Learning update-2}, and \eqref{Eq: Learning update-3}) are required and updated iteratively.

\section{Numerical Results}
\label{Sec:num}

We consider Rayleigh fading with unit variance and the path loss model $68.5+16.1\log_{10} d$ (dB) at the 5.9\,GHz carrier frequency \cite{Liu},
where $d\in[1,100]$ measured in meters is the distance between the camera/MEC server and vehicle.
The channel gain is further quantized into two levels.  There are four cameras in the simulated network.
Considering that the MEC server must have sufficient resources to serve a large number of VUEs, we select $W_{\rm s}=20$\,MHz, $W_{\rm c}=100$\,kHz,  $f_{\rm s}=2\times 10^{11}$\,cycle/s, and 
$f_{i}=10^9$\,cycle/s by referring to \cite{Liu,Han}.
To synthesize images, the same objects in different images need to be recognized first. Therefore, we refer to the processing density for face recognition, i.e., $L=2339$\,cycle/bit \cite{Kwak_JSAC15}, in the simulations.
Some parameters will be specified while investigating their impacts on performance.
The rest simulation parameters are $A=20$\,kbit, $B=60$\,kbit, $N_0=-174$\,dBm/Hz,
 $P_{\rm c}=20$\,dBm, $P_{\max}=30$\,dBm, $\xi=10$, $ \eta_{u}(t)=t^{-0.51}$, $\eta_{r}(t)=t^{-0.52}$, and $\eta_{\pi}(t)=t^{-0.53}$.
We run 10000 iterations in Algorithm \ref{Alg: distributed non-regret}.
For performance comparison, we consider the following four baselines: 
\begin{enumerate}
\item
{\bf Average-based} scheme in which $\mathbb{E}[ T^{\rm E2E}_{i}]$, $\sum_{i\in\mathcal{V}_{\rm o}} T^{\rm DL}_{i}$, and $u_i(t)= -T^{\rm E2E}_{i}(t)$ are considered in \eqref{Eq: VUE_problem_1}, \eqref{Eq: Server_problem}, and \eqref{Eq: Learning update-1}, respectively;
\item 
{\bf Fully-fetching} scheme with $\Pr(\alpha_i=0|\mathbf{h}_{i})=1,\forall\,i\in\mathcal{V},\mathbf{h}_{i}\in\mathcal{H}_i$;
\item
{\bf Fully-offloading} scheme with $\Pr(\alpha_i=1|\mathbf{h}_{i})=1,\forall\,i\in\mathcal{V},\mathbf{h}_{i}\in\mathcal{H}_i$ and $\sum_{i\in\mathcal{V}}\exp(\rho T^{\rm DL}_{i})$ in \eqref{Eq: Server_problem};
 \item
{\bf Half-fetching half-offloading} scheme with $\Pr(\alpha_i=0|\mathbf{h}_{i})=\Pr(\alpha_i=1|\mathbf{h}_{i})=0.5,\forall\,i\in\mathcal{V},\mathbf{h}_{i}\in\mathcal{H}_i$ and $\sum_{i\in\mathcal{V}_{\rm o}}\exp(\rho T^{\rm DL}_{i})$ in \eqref{Eq: Server_problem}.
\end{enumerate}

\begin{figure}[t]
%\vspace{-0.3cm}
\center
\includegraphics[width=\columnwidth]{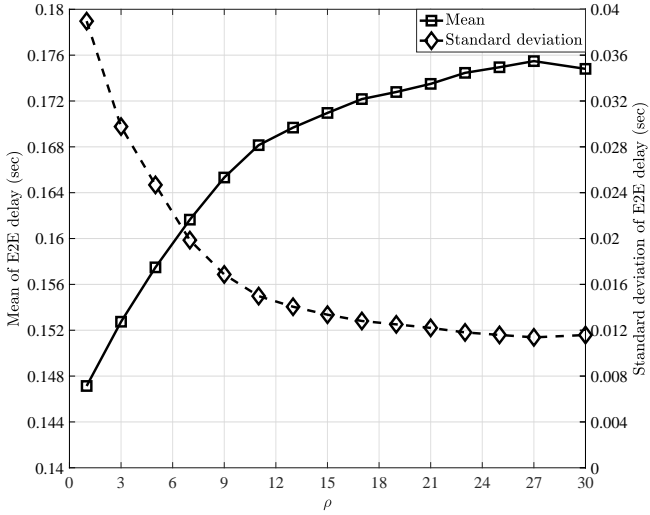}
%\vspace{-0.4cm}
\caption{Mean and standard deviation of the VUE's E2E delay versus $\rho$ with $V=60$ VUEs.}
\label{varying rho}
\end{figure}
\begin{figure}[t]
%\vspace{-0.4cm}
\center
\includegraphics[width=\columnwidth]{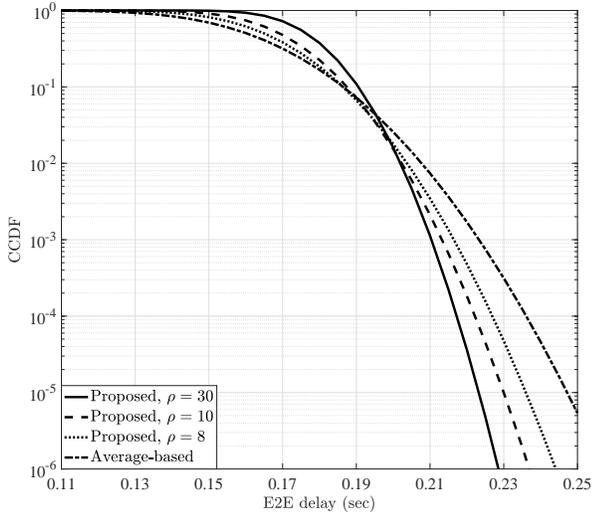}
%\vspace{-0.4cm}
\caption{CCDFs of the VUE's E2E delay for various $\rho$ in our proposed approach and the average-based baseline with $V=60$ VUEs.}
\label{CCDF}
%\vspace{-0.6cm}
\end{figure}

In Fig.~\ref{varying rho}, we show the mean and standard deviation of the E2E delay by varying the risk-sensitive parameter $\rho$. As per the Maclaurin series expansions of the objective functions in \eqref{Eq: entropy_VUE_problem_1} and \eqref{Eq: VUE_problem_1}, the goal is to lower the variance and higher-order statistics which increase with $\rho$. Consequently, the standard deviation (or the variance) of the E2E delay is a decreasing function of $\rho$, whereas the average E2E delay increases with $\rho$. 
Now let us further look into the  the complementary cumulative distribution function (CCDF) of the E2E delay in Fig.~\ref{CCDF}, where we show the CCDFs of the E2E delay of both the proposed and {\bf average-based} approaches.
As shown in the figure, the CCDF of the proposed approach decays faster resulting in a steeper tail, particularly with larger $\rho$.
This behavior means that our approach exhibits lower occurrence probability of extremely high delay, compared to the {\bf average-based} baseline. 
In addition, focusing on the region $[10^{-2},1]$ of the CCDF, we can  see that the proposed approach achieves a lower variance at the expense of higher average performance.

\begin{figure}[t]
%\vspace{-0.3cm}
  \centering
  \subfigure[Mean]
    {\includegraphics[width=\columnwidth]{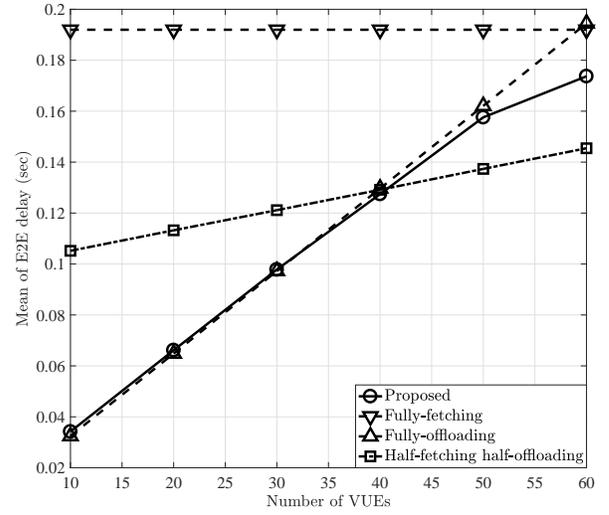}}
 \subfigure[Standard deviation]
   {\includegraphics[width=\columnwidth]{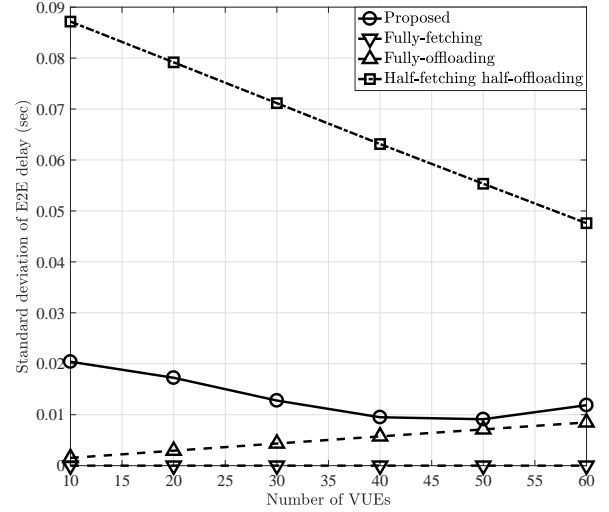}}
    % \vspace{-0.2cm}
  \caption{Mean and standard deviation of the VUE's E2E delay in our proposed approach with $\rho=30$ and the three baselines versus network density.}
  \label{varying VUE}
 % \vspace{-0.7cm}
\end{figure}

Considering different network densities in Fig.~\ref{varying VUE}, we compare the mean and standard deviation of the VUE's E2E delay of our proposed approach and the  three baselines. Since the camera transmits at its lowest rate, and the VUE waits for all cameras' images before synthesizing, the E2E delay in the {\bf fully-fetching} scheme hardly changes, i.e., almost zero variance. However, the average E2E delay in this baseline is the highest due to the camera's lowest transmission rate and the VUE's weak computation capability. When the VUE density is low, more communication and computational resources of the MEC servers are allocated to each VUE, making the {\bf fully-offloading} baseline achieve the lowest average E2E delay. In the low-density regime, our approach gives a solution which is almost equivalent to the {\bf fully-offloading} scheme with the best average performance. In addition, due to the non-zero probability of the task-fetching decision, i.e., $\Pr(\alpha_i=0|\mathbf{h}_{i})\neq 0$, as per \eqref{Eq: no-regret prob}, the delay variance of our proposed approach is slightly higher than the variance of the {\bf fully-offloading} baseline. When the network becomes dense, less resources will be allocated from the MEC server.  In contrast with the {\bf fully-offloading} baseline, our approach increases the probability of the task-fetching decision $\Pr(\alpha_i=0|\mathbf{h}_{i})$ so that the average E2E delay is improved. 
The  superior mean performance of our approach is rather significant compared with the standard deviation degradation.
Finally, due to  equal likelihood for both local computation and task offloading, the VUE in the {\bf half-fetching half-offloading} scheme has the highest delay variance regardless of the network density.

\section{Conclusions}
\label{Sec:con}

In this letter, we have studied the task-fetching and offloading problem in VEC networks towards URLLC.
The studied problem was formulated as a risk minimization problem per VUE's E2E delay in which the entropic risk measure is defined as our reliability metric.
Leveraging a distributed no-regret learning algorithm, we have proposed a risk-sensitive task-fetching and offloading solution. Numerical results have shown a reduction in delay variance  compared to the average-based approach and  other baseline schemes.

\bibliographystyle{IEEEtran}
\bibliography{Ref_VEC}

\end{document}